# Intrinsic Contextuality as the Crux of Consciousness

D. Aerts, J. Broekaert & L. Gabora
Center Leo Apostel, Brussels Free University, Belgium

## 1. THE CONTEXTUALITY OF CONSCIOUS EXPERIENCE

A stream of conscious experience is extremely contextual; it is impacted by sensory stimuli, drives and emotions, and the web of associations that link, directly or indirectly, the subject of experience to other elements of the individual's worldview. The contextuality of one's conscious experience both enhances and constrains the contextuality of one's behavior. Since we cannot know first-hand the conscious experience of another, it is by way of behavioral contextuality that we make judgements about whether or not, and to what extent, a system is conscious. Thus we believe that a deep understanding of contextuality is vital to the study of consciousness.

Methods have been developed for handling contextuality in the microworld of quantum particles. Our goal has been to investigate the extent to which these methods can be used to analyze contextuality in conscious experience. This work is the fledgling efforts of a recently-initiated interdisciplinary collaboration.

## 2. PHYSICAL AND CONCEPTUAL CLOSURE ENABLE CONTEXTUALITY

Most hold to the intuition that inorganic substances are not conscious; that there is nothing it is *like* to be a rock. Even those who believe that inorganic substances *do* have some sort of conscious experience tend to agree that the consciousness of living organisms is an order of magnitude greater. This intuition may stem from the fact that the organism, being a physically closed system, is richly interconnected (e.g., by way of the neural, sensorimotor, and endocrine systems), and this interconnectedness endows it with behavioral contextuality. That is, any perturbation will percolate through the interconnected system and elicit a response tailored to the specifics of the perturbation. Enhanced behavioral contextuality is often viewed as evidence that a system is more conscious.

Humans appear to be endowed with a second level of closure; their memories and sensorimotor associations are interconnected via concepts of varying levels of abstraction into a conceptual web, or worldview. Thus every concept, belief, etc. impacts and is impacted by a sphere of related concepts, beliefs etc., and the network is closed in the sense that there exists a 'conceptual pathway' through streams of associative recall from any one concept to any other.

Gabora [7, 8] suggests that this happens through an autocatalytic process analogous to that proposed by Kauffman [10] to explain the origin of life. The distributed storage and retrieval of memories prompt the emergence of abstractions, and as the density of abstractions increases, the probability they crystalize into an interconnected worldview increases exponentially. For this to happen, the sphere of concepts activated by any perturbation (that is, the extent to which storage/retrieval is distributed) must fall within an intermediate range; the system is poised at the proverbial 'edge of chaos'. A viable and coherent worldview is one that reinforces thought trajectories that enhance wellbeing at the individual and societal levels.

To summarize, physical closure increases the potential for contextuality. Conceptual closure increases it an order of magnitude further, by enabling the individual to engage in relational streams of thought that refine potential actions in light of goals or imagined outcomes. Thus the capacity to give evidence of consciousness is enhanced.

## 3. WEAK VERSUS STRONG CONTEXTUALITY

A large component of conscious experience involves the formation of opinions, i.e. the stabilizing of perceived relationships between self and other. We view opinions as states of the conceptual network in response to an experimental perturbation in the form of a question. Opinions are more stable than some other sorts of conscious experience, such as fleeting perceptions or emotions. However, they too can be affected by context.

We say that an opinion is *weakly contextual* if it seems to have settled into one fairly stable state, or another. For example, you might either love a particular radio station, or dislike it. We say the opinion is *strongly contextual* if it is unstable, prone to sway as easily in one direction as another. This is just an amplitude difference; next we will discuss qualitatively different types of contextuality.

## 4. CLASSICAL VERSUS INTRINSIC CONTEXTUALITY

There are two kinds of contextuality. In *classical contextuality*, the outcome is affected by various aspects of the environment, but not by the irreducible and nonpredictable specifics of the interaction between the system and the experimental perturbation. When the situation is analyzed in terms of states, experiments, and outcomes, Kolmogorov's axioms are satisfied, and a classical probability model can be used. As an example of classical contextuality in a cognitive situation, if you were asked 'do you like this room?' your answer would depend much more on the room than on how the question was asked. This situation is schematically diagrammed in Figure 1.

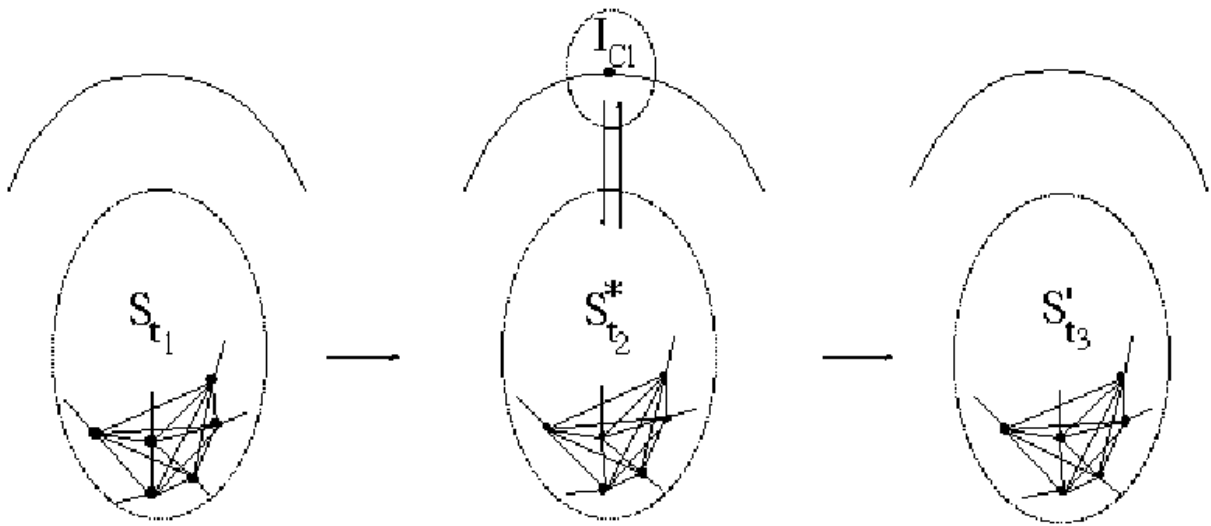

**Figure 1. Schematic representation of a measurement interaction exposing classical contextuality: (t1) The system prior to contextual interaction. (t2) The classically contextual input system reconfigures the initial state and produces adapted response. (t3) The reconfigured system.**

On the other hand, it may be that the outcome is determined through the interaction of the system with the irreducible and nonpredictable properties embedded in the measurement process. This is referred to as *intrinsic contextuality*. The system and the perturbation both have an internal relational constitution, such that their interface creates a *concrescence* of emergent, dynamic patterns. The presence of intrinsic contextuality means that Kolmogorovian axioms are not satisfied, which renders the formal description of the entity nonclassical, or possibly, quantum mechanical-like.

As an example of intrinsic contextuality in cognition, there may be no objective answer to the question of whether you like the radio station or not. Your opinion 'collapses into existence' depending on the irreducible and nonpredictable specifics of the interaction between you and the questioner. Thus, when a friend complains about the radio station, she evokes aspects of it that you do not like, and you may agree with her. But when a smiling representative of the station seductively asks if you like the station, a positive opinion may be manifested. Much as the experiment cannot be performed without employing *some* apparatus, the question cannot be asked without *some* tone of voice, *some* facial expression, etc. This situation is schematically diagrammed in Figure 2.

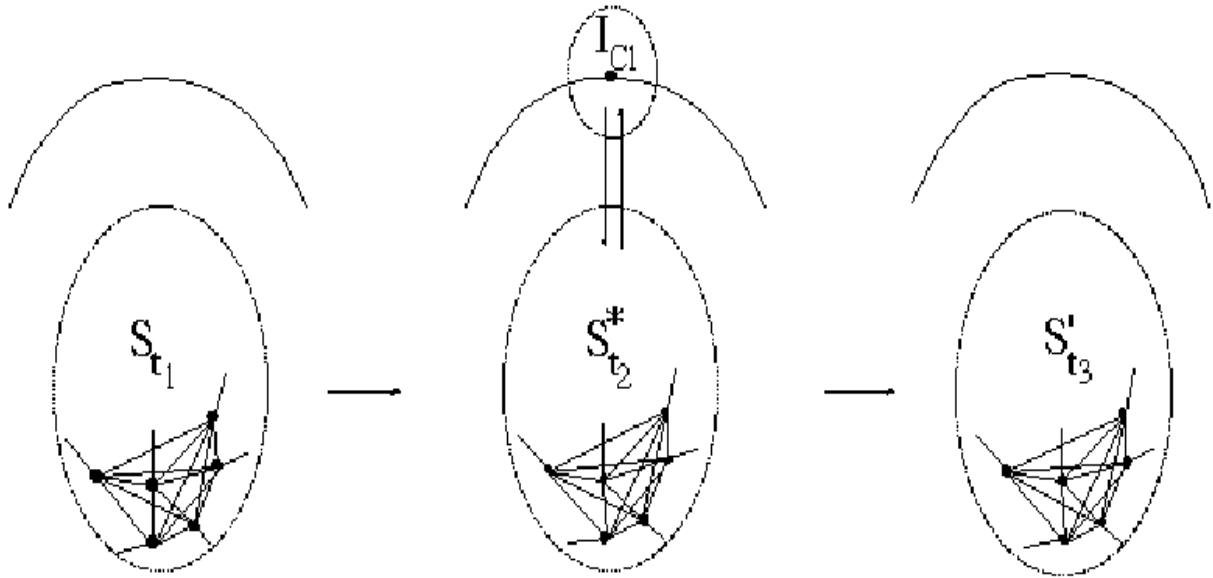

**Figure 2. Schematic representation of a measurement interaction exposing intrinsic contextuality: ($t_1$) The system prior to contextual interaction. ($t_2$) The concrescence of intrinsic contextual input exposes a quantum-like non-predictiveness, symbolically represented here by the superposition "+". ($t_3$) The system collapses into a reconfigured state.**

In practice, all perturbations involve a mixture of classical and intrinsic contextuality in varying degrees.

## 5. A MODEL OF INTRINSIC CONTEXTUALITY

Intrinsic contextuality can be illustrated using a simple model: the quantum machine. The model has been analyzed intensively as a conceptual tool in quantum mechanics [1, 2, 3, and 7] but it is perhaps most useful for explaining quantum-like properties at the macroscopic scale.

The model is defined within a framework of formal model building: an entity is described by means of its set of states, its set of experiments, and the probabilities connected with sets of outcomes. The states of the quantum machine are the different possible places on the surface of a sphere with radius 1 (Fig 3). There is a fairly large variety of states. By placing a point P somewhere else on this surface, another state is realized.

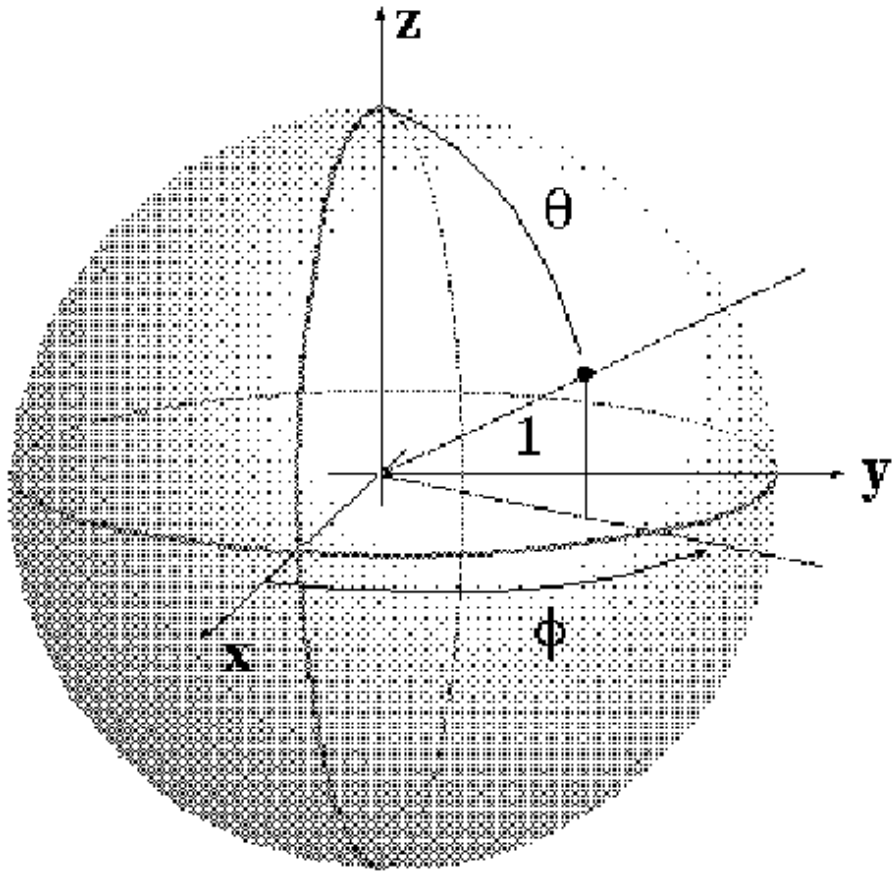

**Fig. 3. The quantum machine: each state is given by a point P on the sphere of radius 1, and can be parameterized by the polar coordinates (q, f).**

In this model one cannot know the location of point P by simply looking at it. We define the set of allowable experiments by which we can find out where P is, which in turn tells us its state. These experiments are made as follows:

- Place an elastic strip (e.g. a rubber band) between two diametrically opposite points of the surface, u and -u (Fig. 4a).

- Point P then falls from its initial place orthogonally onto the elastic, and sticks to it (Fig. 4b).

- Next, the elastic breaks at some arbitrary point, not depending on where the point actually sticks to the elastic. The chance of breakage is evenly spread over the entire length of the elastic.

- Consequently, the point P, attached to one of the two pieces of the elastic (Fig. 4c), is pulled to one of the two endpoints u or -u (Fig. 4d). Depending on where point P ends up, we say the outcome of the experiments is u or -u.

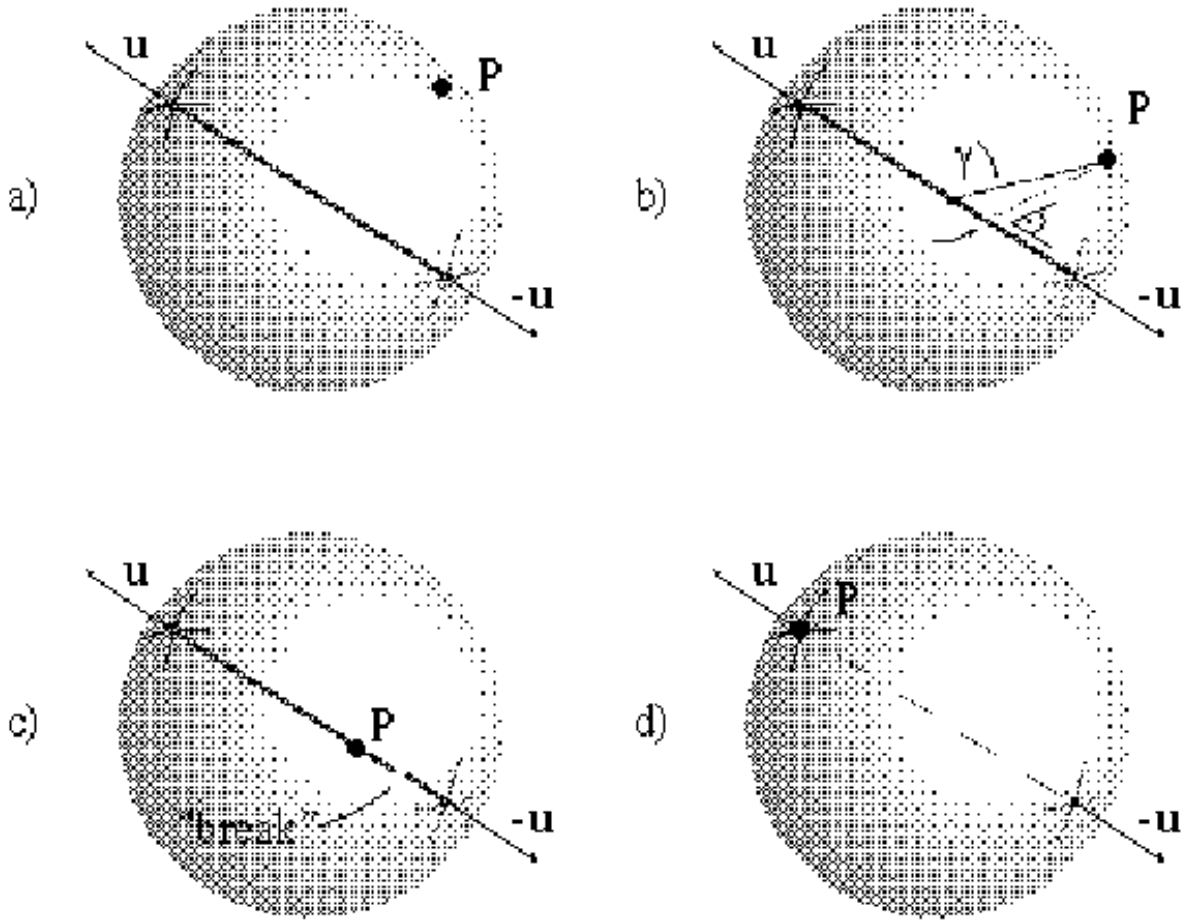

**Fig. 4: The measurement process in the quantum machine model: a) Point P indicates the state. b) It drops orthogonally onto the elastic. c) The elastic breaks at random. d) Point P ends up being pulled to its final state by the residual piece of elastic.**

Why such an intricate way of experimenting on this quantum machine? First, this measurement procedure is set up to discern whether the point P is in some particular state u or -u. We are therefore limited to two possible outcomes. A series of such experiments, varying each time the choice of u, gives a more detailed description of the system.

This model presents a macroscopic situation wherein an inevitable and irreducible effect takes place during the measurement, leading to quantum mechanical outcome probabilities. The effect is irreducible in the sense that it is due to the mere coupling in an interaction between two systems which have different levels of structural organizational composition, causing the probabilistic outcome. (If it were classical contextuality, the measurement effects would be reducible through their determinism, i.e. their effect could be quantitatively described, making it possibile to correct the measurement outcome.)

The probability Pr ( P, u ) (or Pr ( P, -u ) ) of ending up in the point u (or -u) is given by the distance between the point P sticking to the elastic and the opposite point -u (or u) divided by 2, the diameter of the sphere. It can be easily seen that these can be formulated as:

Pr ( P, u ) = $\cos^2 g/2$

Pr ( P, - u ) = $\sin^2 g/2$

It has been proven that any intrinsically contextual entity can be modeled using this quantum machine model (or a variant of it), and that the probabilities that arise are precisely the well known probabilities that describe the spin of a spin-1/2 quantum particle, e.g. the Stern-Gerlach experiment [1, 2, 3, 4, 9].

In the present model, we identify the configuration with the elastic, the process of sticking to it, and its randomized breakage, as intrinsic and inevitable aspects of measuring the state P of the entity. The measurement act indeterministically creates the quantum appearance of outcome state. The interaction of macroscopic systems will, to varying degrees, expose this intrinsic type of contextuality.

## 6. INTRINSIC CONTEXUALITY IN A CONSCIOUS SYSTEM

The above model conveys the basic concept of how an intrinsically contextual interaction can cause a superposition of possible states to collapse into one state or another. We now introduce the Liar paradox as an example of how this arises in conscious processes. (For example, the sentence "This sentence is false.") We have been able to prove that this paradox entails a full quantum mechanical description: the states (truth and falsity) are represented by rays of a complex Hilbert space, the experiments (posing the paradox, asking whether it is true or false) by self-adjoint operators, and the dynamics by a Schroedinger equation [5, 6].

The Liar paradox was chosen because neither state, once settled upon, is acceptable, and so provides the impetus that reiterates the process. Thus we get the opportunity to re-experience the superposition state an arbitrary number of times, which heightens our awareness of it. (Optical illusions can be another example of this.) This kind of superposition is ubiquitous in conscious experience. It is often something intrinsic to the asking of a question, or to the specifics of an accident, that makes you respond one way versus another.

## 7. INCREASING THE INTRINSIC CONTEXUALITY OF A CONSCIOUS SYSTEM

We have seen how formalisms that deal with intrinsic contextuality in the quantum mechanical realm can be adapted to model intrinsic contextuality in a macroscopic system. This is useful because the degree of contextuality of a conceptual system could provide a rough measure of the extent to which it is conscious. We now ask: how does the contextuality of this system increase?

We suggest that the worldview is the subjective experience of the relational structure of the conceptual system, and that structure in the world is gradually incorporated into the worldview through the assimilation of experimental perturbations (in the form of sensory inputs, ideas, etc.). In keeping with Kauffman (1993), we suggest that this is accomplished by oscillating between a *supracritical phase*--wherein the system is robust enough to accept new inputs--and a *subcritical phase*--wherein assimilation of the new inputs temporarily challenges the system's robustness, such that new inputs are not accepted. (The subcritical phase is analogous to annealing in physical systems such as spinglasses.)

With each oscillation, the statespace of the closed conceptual network expands. This expanded statespace in turn increases the capacity of the network to assimilate increasingly complex inputs the next time around; thus there is a coevolutionary relationship between the worldview and the novelty it assimilates.

## 8. CONCLUSIONS

Judgements of the extent to which a system is conscious appear to be related to the *contextuality* of its behavior, which is both enabled and constrained by *closure* in physical space. Contextuality can be further enhanced by closure in conceptual space to form a relationally structured conceptual network or mental model of the world, subjectively experienced as a worldview [7, 8].

Sensory experiences or perturbations, which leave imprints on a cognitive system in the form of novel memories, can be viewed as observations or measurements of the current state of the conceptual network. *Conceptual closure* both facilitates and constrains the long-term assimilation of these memories. *Classical contextuality* can be mathematically corrected for because it arises due to predictable factors of the experimental perturbation. *Intrinsic contextuality* is unavoidable because it arises in virtue of the irreducible and unpredictable perturbation itself. The particular nature of the intrinsically contextual interaction is caused by emergent level relations between the system and the measurement. Since intrinsic contextuality gives rise to quantum structure, our analysis adds to the plausibility that quantum structure plays a role in consciousness.

The closed conceptual network oscillates between a *supracritical phase*--wherein the system is robust enough to accept new inputs--and a *subcritical phase*--wherein assimilation of the new inputs temporarily challenges the system's robustness, such that new inputs are not accepted. We propose that, with repeated oscillations, pattern in the external world is gradually assimilated into the conceptual network, resulting in increased potential for intrinsic contextuality, and increased capacity to express consciousness.